\documentclass{aa501}
\usepackage{graphicx}

\begin{document}

%\thesaurus {06 (04.19.11, 08.02.2, 08.02.4, 08.05.1, 08.06.3)}
\title{GAIA accuracy on radial velocities assessed from a synthetic 
spectra database
}
%\subtitle{}
\author{
Toma\v{z} Zwitter%\inst{1}
}
\offprints{T. Zwitter}
\institute{
University of Ljubljana, Department of Physics, Jadranska 19, 1000 Ljubljana, 
Slovenia \\
%\email
{tomaz.zwitter@uni-lj.si}
}
\date{Received date..............; accepted date................}

\abstract{
Spectrograph aboard the GAIA satellite operates in the near-IR, in 
the 8490-- 8740 \AA\ window accessible also from the ground. 
The most important parameter yet to be determined is the spectral resolution. 
Realistic estimates of the zodiacal light background are obtained and a total 
of $2\times 10^5$ correlation runs are used to study the accuracy of 
radial velocity measured by the spectrograph as a function of resolution, 
magnitude of the target, its spectral type and luminosity class. 
Accuracy better than 2~km/s 
is achievable for bright stars if a high enough dispersion is chosen. 
Radial velocity error of 5 km/s is at $V=17.5$ for Cepheids and 
at 17.7 for horizontal branch stars. Even for very 
faint objects, with spectra dominated by background and readout noise, 
the optimal dispersion is still in the 0.25 /  0.75 \AA/pix range. This 
is also true for complicated cases such as spectroscopic binaries or if 
information other than radial velocity, i.e.\ abundances of individual 
elements or stellar rotation velocity, is sought after.  
The results can be scaled to assess performance of future ground based 
instruments. 
\keywords{Stars:kinematics -- Surveys -- Space vehicles:instruments 
-- Techniques:radial velocities}
}
\maketitle

\section{Introduction}

GAIA is the approved ESA Cornerstone 6 mission designed to obtain 
extremely precise astrometry (in the micro-arcsec regime), multi-band
photometry as well as spectroscopy for up to a billion stars in our Galaxy
and beyond. The goals are described in the mission 
{\it Concept and technology study report}
(ESA-SCI(2000)4, hereafter ESA2000), and summarized by Perryman 
et al.\ (2001).  Astrometric 
information will be used to derive star's position and distance. Proper 
motion then yields the projection of the velocity vector on the sky plane.
The missing, sixth component in the position-velocity space, i.e.\ 
the radial velocity of the target, is derived by spectroscopic observations 
on board (Munari 1999).  Apart from astrometry and spectroscopy a number 
of photometric bands (15 are currently baselined) will collect broad- 
and narrow-band photometric information useful for general 
classification purposes (see Sect.\ 3.3). Spectroscopy can be used to obtain    
additional astrophysical information that cannot be provided by other means. 
This includes abundances of individual elements, stellar 
rotational velocity ($v \sin i$),  and information on line profiles 
which is essential for derivation of orbital solutions of spectroscopic 
binaries (Munari et al.\ 2001a) or for detection and classification of 
peculiar stars (Munari et al.\ 2002). The spectrograph is described 
in ESA2000 and Munari (2001). 

\begin{figure*}
\centering
\includegraphics[width=10.5cm,height=19.4cm,angle=270]{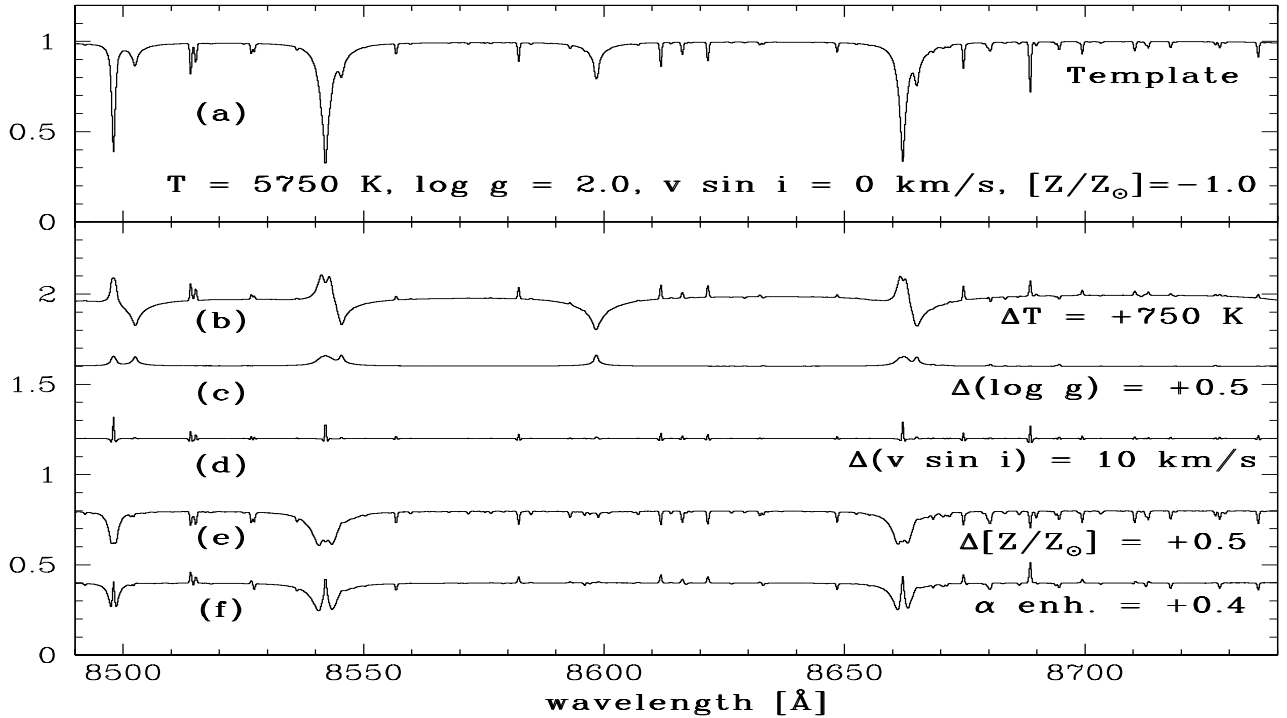}
\caption[]{(a): Synthetic spectrum of a non-rotating G-type 
metal-deficient giant at a $R=20000$ resolution. (b-f):
Spectral change if one of the spectral 
parameters is modified. Each curve is the ratio of the given spectrum 
divided by the top template. The values of the changing parameter are 
indicated. All tracings share the scale of the top panel, 
but are vertically offset for clarity. 
}
\end{figure*}

Performance of the spectrograph can be assessed by convolution of 
spectra of real stars (Munari et al. 2001).
Here we resort to synthetic spectra that give more flexibility to simulate 
parameters of the GAIA spectrograph. The goal of this paper is to assess 
the accuracy of the radial velocities and to some extent other information 
as a function of the chosen spectral dispersion. We upgrade the initial 
assessment (ESA2000) by a database of more realistic spectra, a better treatment 
of the background, better correlation technique, updated spectrograph 
parameters and extending results to significantly fainter magnitudes.
Most important, the initial assessment was done for only one 
resolution  ($R=30000$ sampled at 0.75\AA / pixel, i.e.\ an undersampled case) 
while here we explore a range of possible choices 
($2000 < R < 20000$,  always maintaining the usual criterion of 2 pixels per 
resolution element). Given the fact that the spectral resolution  
is the only spectrograph parameter yet to be decided and the possibility 
to scale the results to any ground-based telescope observing in the 
near-IR, the paper provides realistic error estimates for any future 
campaigns in the near-IR domain. 
   
\section{Synthetic spectra database}

We used Kurucz models to calculate a large database of $\sim 10^5$ 
synthetic spectra that sample the grid in temperature, surface gravity,
metallicity, rotational velocity and spectral resolution. The grid 
which will be described separately (Zwitter et al.\ 2002) upgrades the 
initial calculations of Munari \&\ Castelli (2000) and 
Castelli \&\ Munari (2001). Fig.~1 illustrates the dependence of 
spectral properties on basic model parameters. All tracings are ratios 
to the template spectrum of a non-rotating metal-deficient giant of 
spectral type G, which is shown in Fig.~1a. Plotting the ratio to the 
chosen template instead of the spectrum itself permits a quick 
evaluation of the signal to noise and resolution needed to determine 
the value of a given parameter. Rise of effective temperature weakens 
the Ca~II lines while the Pashen lines get stronger (Fig.~1b). 
Increase of gravity (Fig.~1c) also diminishes the width of the Ca~II lines. 
Stellar rotation (Fig.~1d) has a pronounced effect only for fast rotators 
($v \sin i > 10$~km/s), but the smearing is different from that of 
increase in gravity. The metallicity content (Fig.~1e) influences many lines, 
but details are different from that of enrichment with the $\alpha$-elements
(Fig.~1f). These kind of dependencies are typical for relatively low 
temperature stars that will be the most common GAIA targets. 

In the next section we use few of the calculated spectra to assess 
the performance of the spectrograph for some representative targets.

\section{Simulations}

\subsection{Spectrograph characteristics}

Radial velocity spectrometer instrument aboard GAIA is a slitless 
spectrograph operating in a  time delay integration mode. The field of view is 
$2^o \times 1^o$. The telescope has a rectangular aperture of 
$0.75 \mathrm{m} \times 0.7\mathrm{m}$. The focal plane is covered 
with an array of CCDs. The total efficiency of the telescope + 
spectrograph + CCD is baselined to 0.35. Each star drifts across the 
plane in 60.4~s, with its signal read by two consecutive CCDs. The CCDs 
have 20~$\mu$m pixels, corresponding to 1 arcsec on the sky. The 
readout noise of each CCD is 3.0 e$^-$. The dispersion direction is 
perpendicular to the drift. Each star has all signal on one or two 
pixels in the drift direction.  It was suggested to study a 
possibility of binning together the pixels 
in the drift direction for faint objects and so avoid excessive RON. 
The background is calculated from a suitable region displaced along 
the direction of the drift scan. Altogether the number of pixels 
sent to the Earth is $3 \times 0.06 R$ for bright stars, and 
$2 \times 0.06 R$ for faint stars binned across the spatial direction
(if the binning is used). 

\subsection{Noise sources}

%\begin{figure}[!t]
\begin{figure}
\includegraphics[width=6.7cm,angle=270]{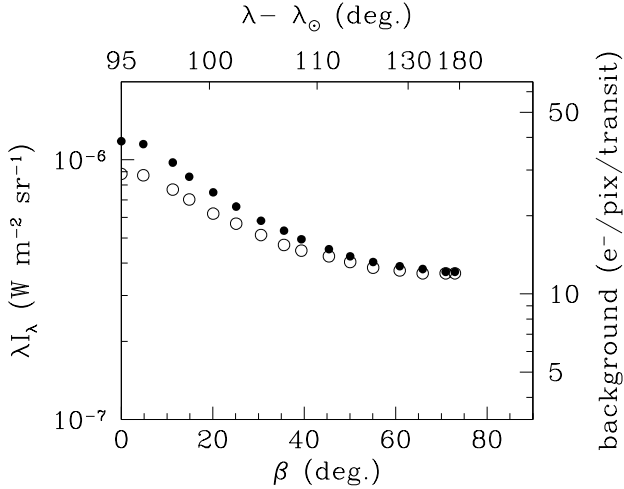}
\caption[]{
Ecliptic latitude dependence of the zodiacal background for the 0.5~$\mu$m
(filled symbols) and 1.43~$\mu$m (open symbols) bands. Adapted from 
Matsumoto et al.\ (1996). The scale on the right gives the corresponding 
number of electrons per pixel per observation detected by GAIA. 
}
\end{figure}

%\begin{figure}[!t]
\begin{figure}
\includegraphics[width=8.8cm,angle=0]{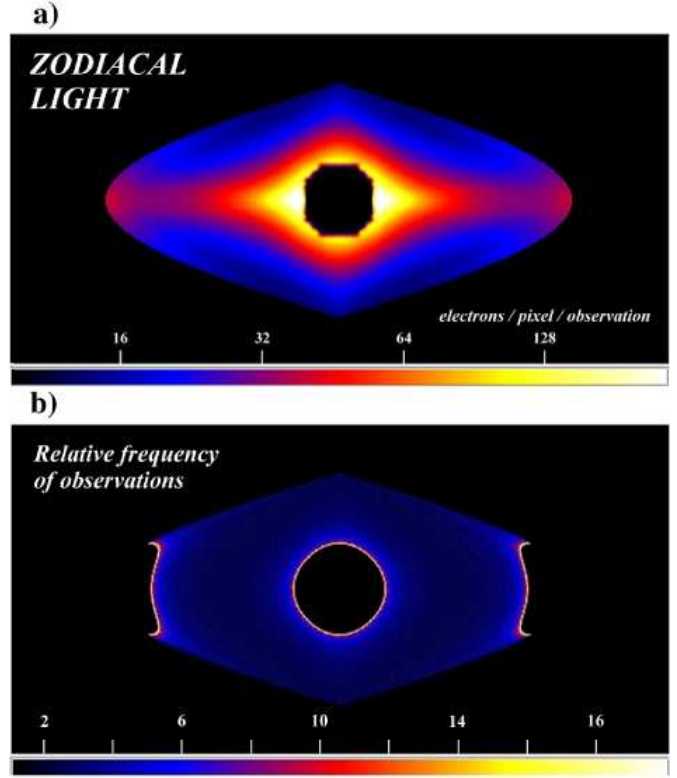}
\caption[]{(a) Number of electrons per pixel per transit due to zodiacal 
light background. The data are shown in  helioecliptic coordinates, where 
the Sun stays at the center of the figure and equator coincides with the ecliptic. 
(b) Relative number of transits in the same coordinates.
}
\end{figure}

For bright targets the results will be limited by Poisson noise, but at 
the faint limit the read-out noise and noise of the background become more 
important. GAIA will operate from the second Lagrangian point of the Sun-Earth 
system. Save for the densest sky regions the dominant background signal 
will be zodiacal light (ESA2000, see also Sect.\ 5 below). Here we explore its spatial variation 
in some detail.

The zodiacal background observed by GAIA will be much the same as seen 
from Earth, as the cloud of scattering particles extends well beyond 
the GAIA orbit (Kelsall et al.\ 1998, Gorkavyi et al.\ 2000). Observations 
of the Infrared Telescope in Space (Matsumoto et al.\ 1996) show that sky 
brightness at 0.5 and at 1.43 $\mu$m has very similar 
energy density and similar dependence on the viewing direction
(Fig.~2). This is consistent with earlier balloon observations 
at 0.71 $\mu$m and 0.82 $\mu$m (Frey et al.\ 1974). The angular 
distribution of the zodiacal brightness is however remarkably un-uniform. 
It depends on ecliptic coordinates of the viewing direction
(here we neglect minor asymmetries vs. ecliptic plane, see 
James et al.\ 1997). Fig.~3a shows the number of zodiacal 
light photons ($N_{sky}$) per pixel detected during a single 60.4~s transit.
The number is calculated from data in Levasseur-Regourd \&\ Dumont (1980)
using the parameters of the spectrograph cited above and assuming that 
a suitable filter blocks away any light outside the GAIA spectral window. 
The numbers show that the zodiacal background noise ($\sqrt{N_{sky}}$) is 
larger than the readout noise for two readouts ($3.0 \sqrt{2} \mathrm{e}^-$), 
except near the ecliptic poles where the two contributions are comparable. 

GAIA spectroscope always points perpendicularly to the axis of the 
satellite, which is aimed at the angle of $55^o$ to the Sun. So the sky is 
not sampled uniformly. Fig.~3b shows that directions at the angle of 
$90^o \pm 55^o$ to the Sun are sampled more often than the ones in between. 
As the telescope circles the Sun each star will have its  helio-ecliptic 
longitude drifting across the whole range, but its ecliptic latitude  
remains constant.

Note that the readout noise and the zodiacal background (per pixel) do not 
depend on the chosen spectral resolution, while the stellar signal per pixel 
is inversely proportional to the spectral resolution. So for faint targets
the background and read-out noise degrade the results more if the spectral 
resolution is increased. 

\subsection{Single transit and mission average spectra}

We ran $2.4 \times 10^5$ simulations to evaluate performance of the GAIA 
spectrograph in measuring stellar radial velocities. Three cases were 
considered: a relatively hot A8 type main sequence star, a cooler K1/2 main 
sequence star and a G6 type supergiant. These choices are representative 
of the most important GAIA targets: a hot relatively luminous star that 
can be seen up to galactocentric distance, a cool dwarf of most common 
type observed by GAIA and a luminous supergiant that can be seen even 
in other galaxies. 

A spectrum of the object was degraded to a desired spectral resolution and 
assigned the flux corresponding to its V magnitude. The spectrum was then 
Doppler shifted to a randomly chosen velocity in the $\pm 100$~km/s range. 
After addition of the zodiacal background the Poisson noise and the readout 
noise (corresponding to two consecutive CCD readouts during a single transit) 
were added. The radial velocity was recovered by correlation with a 
synthetic spectrum (Tonry \&\ Davis 1979). Correlations were done with the 
{\sl xcsao} task within the IRAF's {\sl rvsao} package 
 (Kurtz et al.\ 1992). Degradation of 
spectral resolution, rebinning, Doppler shifts etc. were performed with a 
custom IRAF script.

It would be unphysical to try to recover the radial velocity using the same 
template as was used for initial input. GAIA photometric observations will 
provide quite an accurate guess on the stellar parameters of the target, 
but the match will never be perfect. Typical errors after the end of the 
mission will be 125~K in temperature, 0.4 in 
$[\mathrm{Z}/\mathrm{Z}_\odot]$ and 0.25 in $\log g$ (ESA2000). 
To stay on the conservative side and to allow for 
a limited ability to reproduce observed spectra by theoretical calculation 
we used a larger mismatch: the spectrum used to recover the radial velocity 
had a random mismatch of 250~K in temperature, 0.5 in 
$[\mathrm{Z}/\mathrm{Z}_\odot]$ and 0.5 in $\log g$ . The rotational velocity 
was also different: the initial input spectrum had $v \sin i = 20 $~km/s 
(for A8 V and K1/2 V star), and $2 $~km/s (G6 I star), while the one used 
as a correlation template  had $v \sin i = 10 $~km/s 
(for A8 V and K1/2 V star), and $5 $~km/s (G6 I star).

\begin{table*}
\caption{Radial velocity error ( in km s$^{-1}$) for a background of 74~e$^-$ per 
wavelength bin, typical for an object in the ecliptic plane. Values are given for mission 
averaged spectra (100 transits) while those in
brackets are for a single transit spectrum.}
\begin{tabular}{llrrrrrrr}
Sp.\ Type&Dispersion& R&\multicolumn{6}{c}{Visual magnitude}\\
         & \AA/pix  & $\lambda / \Delta \lambda$&14.0& 15.0 & 16.0 & 17.0 & 18.0 & 19.0\\ \hline \hline
A8 V     &    0.25   &17200& 1.58( 5.43)   &  1.90(24.0 )   &  3.06($>40$)	&  7.84($>40$)  & $>40$($>40$)  & $>40$($>40$)\\
	 &    0.50   & 8610& 2.80( 6.31)   &  3.01(17.0 )   &  3.79($>40$)	&  6.62($>40$)  & 32.2 ($>40$)  & $>40$($>40$)\\
	 &    0.75   & 5740& 4.66( 7.92)   &  4.69(17.7 )   &  5.15($>40$)	&  7.72($>40$)  & 26.0 ($>40$)  & $>40$($>40$)\\
	 &    1.00   & 4305& 6.81(10.9 )   &  6.87(16.6 )   &  7.15($>40$)	&  9.69($>40$)  & 22.1 ($>40$)  & $>40$($>40$)\\
	 &    2.00   & 2160& 13.2(16.5 )   & 13.4 (24.5 )   & 13.9 ($>40$)	& 15.9 ($>40$)  & 25.6 ($>40$)  & $>40$($>40$)\\
\hline	 
K1 V     &    0.25   &17200&  0.23( 1.84)   &  0.40( 4.17)   &  0.92(14.9 )   &  2.30($>40$)   &  5.59($>40$)   & $>40$($>40$)\\
	 &    0.50   & 8610&  0.28( 2.10)   &  0.44( 4.38)   &  0.96( 9.97)   &  2.10($>40$)   &  5.24($>40$)   & 19.4 ($>40$)\\
	 &    0.75   & 5740&  0.29( 2.53)   &  0.47( 5.06)   &  1.08(10.5 )   &  2.23($>40$)   &  5.45($>40$)   & 13.8 ($>40$)\\
	 &    1.00   & 4305&  0.31( 3.00)   &  0.56( 5.50)   &  1.11(12.5 )   &  2.54(39.7 )   &  5.72($>40$)   & 15.3 ($>40$)\\
	 &    2.00   & 2160&  0.57( 5.20)   &  0.93( 9.52)   &  1.67(17.3 )   &  3.35(39.1 )   &  7.18($>40$)   & 19.1 ($>40$)\\
\hline
G6 I     &    0.25   &17200&  0.11( 1.09)   &  0.20( 2.41)   &  0.44( 5.92)   &  1.09($>40$)   &  2.86($>40$)	& 25.9 ($>40$)\\
	 &    0.50   & 8610&  0.17( 1.52)   &  0.31( 3.37)   &  0.65( 7.41)   &  1.52(32.7 )   &  3.48($>40$)	& 12.4 ($>40$)\\
	 &    0.75   & 5740&  0.48( 1.88)   &  0.55( 3.60)   &  0.82( 7.71)   &  1.72(27.2 )   &  4.11($>40$)	& 11.9 ($>40$)\\
	 &    1.00   & 4305&  0.67( 2.28)   &  0.77( 4.25)   &  1.07( 8.31)   &  1.98(23.7 )   &  4.54($>40$)	& 11.4 ($>40$)\\
	 &    2.00   & 2160&  0.63( 3.55)   &  0.77( 6.45)   &  1.25(11.4 )   &  2.74(28.2 )   &  5.69($>40$)	& 13.5 ($>40$)\\
\hline
Cepheid  &    0.25   &17200&  0.34( 2.60)   &  0.59( 5.09)   &  1.28(22.3 )   &  3.14($>40$)   &  7.48($>40$)	& $>40$($>40$)\\
	 &    0.50   & 8610&  0.38( 2.86)   &  0.60( 5.34)   &  1.15(13.4 )   &  2.70($>40$)   &  6.52($>40$)	& 24.6 ($>40$)\\
	 &    0.75   & 5740&  0.63( 3.34)   &  0.78( 5.64)   &  1.32(12.2 )   &  2.85($>40$)   &  6.56($>40$)	& 18.6 ($>40$)\\
	 &    1.00   & 4305&  0.71( 3.32)   &  0.87( 6.77)   &  1.34(13.2 )   &  2.98(31.9 )   &  6.58($>40$)	& 18.1 ($>40$)\\
	 &    2.00   & 2160&  0.63( 5.24)   &  0.97( 9.09)   &  1.86(17.0 )    &  3.54(35.0 )   &  7.50($>40$)	& 19.8 ($>40$)\\
\hline
HB       &    0.25   &17200&  0.34( 2.19)   &  0.52( 4.48)   &  1.07(20.8 )   &  2.51($>40$)   &  7.30($>40$)	& $>40$($>40$)\\
	 &    0.50   & 8610&  0.57( 2.51)   &  0.72( 5.32)   &  1.21(16.4 )   &  2.64($>40$)   &  6.00($>40$)	& 33.8 ($>40$)\\
	 &    0.75   & 5740&  0.90( 3.09)   &  1.00( 5.95)   &  1.47(13.2 )   &  2.75($>40$)   &  6.44($>40$)	& 21.6 ($>40$)\\
	 &    1.00   & 4305&  1.35( 3.98)   &  1.45( 6.24)   &  1.77(14.0 )   &  2.95($>40$)   &  6.66($>40$)	& 23.8 ($>40$)\\
	 &    2.00   & 2160&  2.32( 7.28)   &  2.52(10.2 )   &  2.93(20.9 )   &  4.63($>40$)   &  9.47($>40$)	& 22.8 ($>40$)\\
\hline
\end{tabular}
\end{table*}

\begin{table*}
\caption{Radial velocity error ( in km s$^{-1}$) for a background of 20~e$^-$ per 
wavelength bin, typical for an object at ecliptic latitude $\sim 45^\mathrm{o}$. Values are given for mission averaged spectra (100 transits) while those in
brackets are for a single transit spectrum.}
\begin{tabular}{llrrrrrrr}
Sp.\ Type&Dispersion& R&\multicolumn{6}{c}{Visual magnitude}\\
         & \AA/pix  & $\lambda / \Delta \lambda$&14.0& 15.0 & 16.0 & 17.0 & 18.0 & 19.0\\ \hline \hline
A8 V     &    0.25   &17200&   1.64( 4.63)  &	1.67(15.4 )  &   2.13($>40$)  &   3.98($>40$)  &  14.2 ($>40$)  &  $>40$($>40$)\\
	 &    0.50   & 8610&   2.82( 5.60)  &	2.82(11.0 )  &   3.21($>40$)  &   4.16($>40$)  &   8.98($>40$)  &  $>40$($>40$)\\
	 &    0.75   & 5740&   4.64( 7.54)  &	4.65(11.8 )  &   4.96(35.8 )  &   5.73($>40$)  &   9.57($>40$)  &  $>40$($>40$)\\
	 &    1.00   & 4305&   6.78(10.1 )  &	6.76(13.9 )  &   6.92(32.7 )  &   7.73($>40$)  &  12.4 ($>40$)  &  31.2 ($>40$)\\
	 &    2.00   & 2160&  13.2 (18.0 )  &  13.4 (22.9 )  &  13.6 (39.9 )  &  14.4 ($>40$)  &  18.0 ($>40$)  &  36.0 ($>40$)\\
\hline	 
K1 V     &    0.25   &17200&   0.19( 1.52)  &	0.28( 3.01)  &   0.53( 6.82)  &   1.19(39.4 )  &   2.86($>40$)  &   8.26($>40$)\\
	 &    0.50   & 8610&   0.25( 1.83)  &	0.36( 3.32)  &   0.61( 7.10)  &   1.24(21.9 )  &   2.92($>40$)  &   7.44($>40$)\\
	 &    0.75   & 5740&   0.26( 2.32)  &	0.42( 4.11)  &   0.74( 7.58)  &   1.32(20.2 )  &   3.15($>40$)  &   7.37($>40$)\\
	 &    1.00   & 4305&   0.27( 2.98)  &	0.45( 5.23)  &   0.86( 8.81)  &   1.66(28.6 )  &   3.38($>40$)  &   7.93($>40$)\\
	 &    2.00   & 2160&   0.60( 4.85)  &	0.80( 8.52)  &   1.38(13.5 )  &   2.42(27.2 )  &   5.22($>40$)  &  10.4 ($>40$)\\
\hline
G6 I     &    0.25   &17200&   0.10( 0.98)  &	0.14( 1.86)  &   0.28( 4.05)  &   0.59(26.4 )  &   1.50($>40$)  &   3.96($>40$)\\
	 &    0.50   & 8610&   0.15( 1.38)  &	0.24( 2.36)  &   0.50( 4.78)  &   0.95(14.3 )  &   2.06($>40$)  &   4.54($>40$)\\
	 &    0.75   & 5740&   0.47( 1.70)  &	0.52( 3.02)  &   0.67( 5.79)  &   1.07(14.9 )  &   2.51($>40$)  &   5.58($>40$)\\
	 &    1.00   & 4305&   0.69( 2.11)  &	0.71( 3.60)  &   0.85( 6.77)  &   1.36(14.7 )  &   2.77($>40$)  &   6.08($>40$)\\
	 &    2.00   & 2160&   0.59( 3.79)  &	0.75( 5.72)  &   1.05(10.1 )  &   1.76(19.2 )  &   3.34($>40$)  &   7.85($>40$)\\
\hline
Cepheid  &    0.25   &17200&   0.31( 2.16)  &	0.48( 4.09)  &   0.90( 9.37)  &   2.05($>40$)  &   4.87($>40$)  &  16.6 ($>40$)\\
	 &    0.50   & 8610&   0.36( 2.47)  &	0.48( 4.59)  &   0.89( 8.79)  &   1.80(28.0 )  &   4.18($>40$)  &  11.1 ($>40$)\\
	 &    0.75   & 5740&   0.59( 2.72)  &	0.72( 4.80)  &   1.04(10.1 )  &   1.94(21.2 )  &   4.40($>40$)  &  11.5 ($>40$)\\
	 &    1.00   & 4305&   0.69( 3.55)  &	0.81( 5.81)  &   1.17( 9.89)  &   2.00(22.9 )  &   4.41($>40$)  &  11.1 ($>40$)\\
	 &    2.00   & 2160&   0.58( 5.49)  &	0.95( 8.32)  &   1.59(13.9 )  &   2.85(28.5 )  &   5.28($>40$)  &  11.6 ($>40$)\\
\hline
HB       &    0.25   &17200&   0.31( 1.75)  &	0.44( 3.70)  &   0.79( 8.97)  &   1.68($>40$)  &   4.06($>40$)  &  23.0 ($>40$)\\
	 &    0.50   & 8610&   0.52( 2.31)  &	0.64( 4.22)  &   0.92( 8.69)  &   1.81(37.5 )  &   4.27($>40$)  &  12.1 ($>40$)\\
	 &    0.75   & 5740&   0.88( 3.13)  &	1.03( 4.79)  &   1.22( 9.76)  &   2.02(36.0 )  &   4.67($>40$)  &  11.1 ($>40$)\\
	 &    1.00   & 4305&   1.38( 3.78)  &	1.41( 6.16)  &   1.65(10.2 )  &   2.45(28.2 )  &   4.81($>40$)  &  11.6 ($>40$)\\
	 &    2.00   & 2160&   2.37( 6.89)  &	2.46(10.1 )  &   2.92(17.8 )  &   4.04(37.7 )  &   7.08($>40$)  &  15.9 ($>40$)\\
\hline
\end{tabular}
\end{table*}

Apart from a spectrum obtained during a single passage over the target 
we also simulated mission-averaged spectra. On the average each target 
will be observed  100 times during the  4-year mission. So we co-added 
 100 single mission spectra in order to simulate the ability of deriving 
radial velocities for single non-varying stars. The conditions for such 
summation will be met: each spectrum (and not only the correlation function) 
will be sent to the Earth, and previewed accuracy of wavelength 
calibration will permit their co-addition at the end of the mission.

\section{Results}

Results of the achievable radial velocity accuracy are presented in 
Fig.~4
and Tables 1 and 2. Each point presents a standard radial velocity error, 
$[ \sum_{\mathrm{i=1}}^{\mathrm{N}}{(v_i (\mathrm{true})-v_i (\mathrm{recovered}))^2} /N]^{0.5}$,
as resulting from some $N=400$ tries using a given spectral 
resolution for a star of a given spectral type and magnitude. 
The distribution of errors of each point turns out to be Gaussian 
within statistical errors also for the faintest stars and for all resolutions. 
The {\it xcsao} correlation routine had no appreciable convergence problems 
even for the faintest stars sampled at the lowest spectral resolution (the 
convergence was always succesfull in $>95$\%\ of the tries).  
We used 5 different spectral dispersions, colour 
coded in Fig.~4. The resolution element had always a size of two
pixels. The graphs are in three rows, each for a different type 
of a star. The left column gives results for a single transit spectrum, 
while the right one is for a mission-averaged spectrum. Top labels at 
each graph quote a representative distance of the star, assuming no 
interstellar absorption is present.  

Simulations presented in Fig.~4 assume the background of 74 e$^-$ per 
wavelength bin. This is an average value of zodiacal light background
for an object on the ecliptic plane with the spectrum falling on a 
single pixel row (Fig.~3a) and corresponds 
to $V=21.53/\mathrm{arcsec}^2$. We also run an identical set of 
simulations for a lower value of the sky background
(the adopted value of 20 e$^-$ 
per wavelength bin is an average value for an object $45^o$ away from the 
ecliptic).

The results for the three types of stars are remarkably different. Spectrum 
of a hot A8~V type star (T=7500 K, $[Z/Z_\odot]=0.0$, $\log(g)=4.5$) 
has relatively few lines, so its results are worse than those obtained for 
cooler stars. It is dominated by weak and sharp Ca~II lines 
and the Paschen series lines.  The relative intensity of the Ca~II and
Paschen lines rapidly changes with temperature so a mismatch in the spectral 
type of the star introduces a systematic error in the strength of both sets 
of lines. The correlation routine tries to compensate for line intensity 
mismatch by introducing an additional velocity shift that causes a systematic 
offset in the recovered velocity. This error can be corrected by obtaining 
spectra with better resolution: better sampling of line profiles means 
more points carrying radial velocity information, so that the correlation 
routine can avoid systematic errors in the recovered radial velocity. 
This is well seen in top-right panel of Fig.~4, where the accuracy of 
bright star ($V \leq 15$) measurements depends only on spectral resolution 
and not on the magnitude. The situation for  
faint targets ($V \geq 17$) is different: the adopted relatively 
high background starts to dominate the statistics for the highest resolutions, 
so dispersions of 0.5~\AA/pix  or 0.75~\AA/pix perform best for faint 
hot targets. Note that the point of crossing, i.e. the limit where 
choosing coarser dispersion starts to pay off, depends on the adopted value 
of the background. For another set of simulations with assumed background 
value of 20~e$^-$ (Table~2) the crossing occurs only at magnitude $V=18.0$.

\begin{figure*}
\centering
\includegraphics[width=15.5cm,height=19.5cm,angle=0]{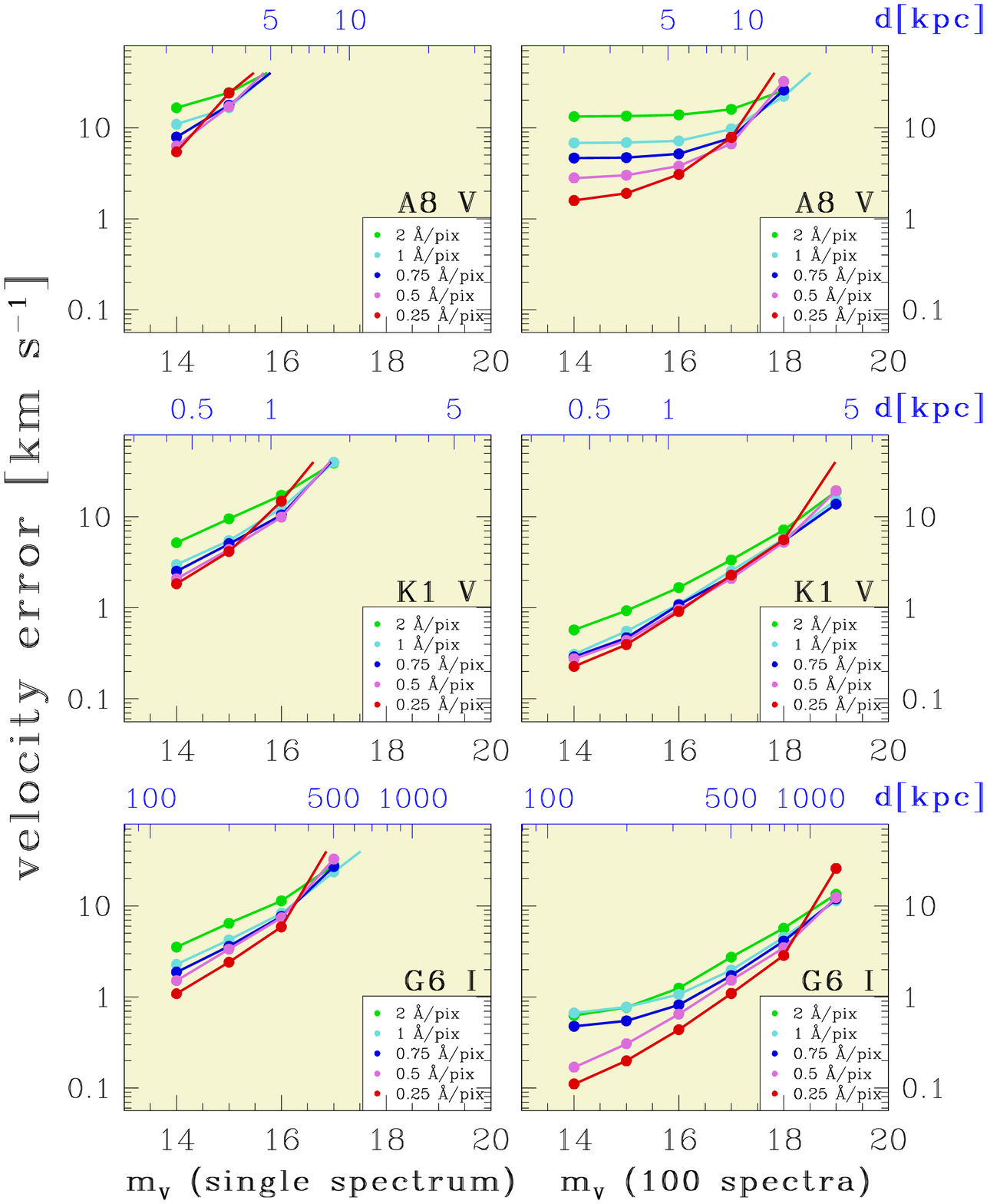}
\caption[]{Average radial velocity determination error as a function 
of stellar magnitude, spectral type and the used spectral dispersion.
Top labels at each graph quote a representative 
distance if no interstellar absorption is present. Left column is 
for single transit spectra, while the right one is for a mission-averaged 
spectrum (100 transits). The assumed background is 74~e$^-$ per 
wavelength bin.  
}
\end{figure*}

Cool dwarfs (K1/2~V: T=5000 K, $[Z/Z_\odot]=0.0$, $\log(g)=4.5$) 
perform better. The reason are very strong, 
core-saturated Ca~II lines that dominate the spectrum. These lines enable 
correlation at fainter magnitudes, so improving the accuracy. And because 
Ca~II lines are very broad and symmetrical the issue of resolution is 
less important. Actually the radial velocity accuracy is very similar 
for all dispersions except for the coarsest one (2~\AA/pix) where 
even the broad Ca~II lines get undersampled. The crossing of the 
0.25~\AA/pix dispersion occurs at $V\sim 15.5$ (spectrum from a single transit) 
and $V \sim 18$ (mission average). If the background were lower 
(20~e$^-$, Table~2) the errors at the faint end would be 50\%\ lower and the 
crossing less pronounced. The results for a mission average 
spectrum are very similar to those of $\sim 2.5$ magnitude brighter stars 
observed during a single transit. If the spectrum had no background and  
readout noise limits, the gain would be a factor of  100 in flux or
a difference of  5 magnitudes. 

Cool supergiants (G6~I: T=4750 K, $[Z/Z_\odot]=0.0$, $\log(g)=1.0$) 
will be optimal targets for deriving accurate radial 
velocities at extreme distances. Their spectra are much richer than 
those of dwarfs of similar temperature, as the Paschen lines and many 
lines of metals start to appear. 
More lines mean more points that can constrain the position of the 
peak of the correlation function. As a result the radial velocity 
is derived with remarkable accuracy. The question of resolution is 
important: better resolution allows for distinction of the Ca~II and Pashen 
lines, as well as deblending of close metallic lines. The crossing 
of the 0.25~\AA/pix dispersion occurs at $V \sim 16.3$ for single 
transit spectra, 
while the highest resolution still performs well even at magnitude $V=18$ for 
a mission averaged spectrum. The reason is the richness of lines so that 
co-addition of many well-sampled spectra is still advantageous. On the other 
hand weak lines in faint stars get lost in the noise, so the question of a 
chosen spectral resolution is less critical at the faint end than it is 
for bright stars. In the case of low background (20 e$^-$) the errors for
the faint stars are 50\%\ lower.

Fig.~4 demonstrates that radial velocities with an accuracy of 
2~km/s in A8 dwarfs are obtainable only for the case of the highest 
spectral resolution (0.25~\AA/pix). These stars can be observed up to 
the distance of $\sim 5$~kpc. For cooler dwarfs (K1/2 V) 
the same accuracy is achievable for fainter targets ($V \leq 17$),
corresponding to smaller distances ($\sim 2$~kpc). The issue of 
spectral dispersion is less important than for hot dwarfs unless the worst 
sampling (2 \AA/pix) is chosen. Note that accuracy achievable from single 
transit spectra is important: most spectroscopic binaries or intrinsically 
variable (spot) stars will be cool dwarfs. Fig.~4 shows that GAIA will 
be able to measure their radial velocity down to $V=15$ with an error 
of $\sim 3$~km/s. But the superposition of various dispersions
can be misleading in this case: spectroscopic binaries or spotty stars 
require a sufficient spectral resolution for any meaningful measurement
(Munari et al.\ 2001a, 2002). The cool supergiants can be used to measure 
kinematics at the largest distances, reaching nearby galaxies.

Tables 1 and 2 give detailed results for two levels of the background.  
 Relative uncertainties of the quoted standard velocity errors are 
estimated at 3\% , as can be seen from comparison of values for bright stars 
in both tables.  
We added results for classical Cepheids 
(T=5500 K, $[Z/Z_\odot]=-0.2$, $\log(g)=2.0$, $v \sin i = 10 $~km~s$^{-1}$) 
and horizontal branch stars 
 (T=5000 K, $[Z/Z_\odot]=-0.5$, $\log(g)=3.5$, $v \sin i = 10 $~km~s$^{-1}$) 
that are important kinematic tracers in 
the Galaxy (ESA2000). 
Results can be readily scaled to any ground based 
spectrograph with collecting area different from 0.53~m$^2$
as the single transit spectra have an effective exposure time 
of 1 minute and the mission averaged ones  100 minutes. 

\section{Discussion}

The results presented in Fig.~4 are compatible with initial performance 
assessment of the GAIA radial velocity spectrometer (Katz 2000). 
Recently Munari et al. (2001) published an extensive 
series of cross-correlations using an observed set of Echelle spectra. 
They limited themselves to stars brighter than those considered here.
But bright stars of identical spectral type are hard to find. So they had 
to use a large spread of spectral types, and the actual mismatch 
between spectra and templates was larger than will 
be for the mission. Correspondingly they derived somewhat larger radial 
velocity errors (within a factor of 2). But they came to the same basic 
conclusion: a high-enough spectral dispersion is a critical factor 
determining radial velocity accuracy for bright stars. 

Here we extend this work to fainter magnitudes. We explore a range of 
spectral types and include simulations of very faint stars having 
a signal to noise ratio for a single transit $S/N < 1$. At such extreme 
conditions a spectrum with the dispersion of 0.25~\AA/pix is limited by 
the sky background. And because this background is mostly zodiacal light
a somewhat lower dispersion (0.5~\AA/pix) performs better for the faintest
($V\ge 18$) dwarfs.  The resolution of 0.75~\AA/pix is also acceptable
for the faintest stars.  

There are other reasons to collect high dispersion spectra. Fig.~1 shows 
that values of all spectral parameters can be accounted for by spectroscopy, 
but note that a high spectral resolution is needed to make secure 
distinctions. All tracings are plot for $R=20000$; halving the resolution 
already mixes up gravity and rotation or metallicity and $\alpha$-element 
enrichment. Additional set of simulations shows that $\alpha$-element 
enrichment of $+0.4$ can be detected down to $V=14$, and abundancies of 
individual elements can be measured for bright stars. Abundance of a 
particularly difficult one, Manganese (with the only useful line 
in the GAIA spectral range at $\lambda =8740.9$~\AA\ with 
$EW \approx 0.02$~\AA ) can be determined within
a factor of 2 for $V<8$. Other elements, i.e.\ N, Si, Fe, Mg and Ti cause 
much stronger lines and so can be studied down to $V \approx 13$.
Stellar rotation is easy to determine for fast 
rotators (Munari 1999), but cool dwarfs commonly observed
by GAIA rotate slowly. Stellar rotation with an error of 10~km/s 
of a $T=5000$~K dwarf can be 
measured for $V<9$. All results quoted here are for the dispersion of 
0.25~\AA/pix and depend critically on the spectral resolution. 

GAIA will observe many spectroscopic binaries. Munari et al. (2001a) showed 
that good quality orbital solutions, including masses, can be derived. 
But also this result critically depends on the resolution of the spectrum. 
Single Angstrom corresponds to 35~km/s at the wavelengths of the GAIA 
spectrograph. Velocity amplitudes of most spectroscopic binaries are 
not much larger than this, so it is clear that a dispersion 
of 0.25 or 0.5~\AA/pix is needed for a proper deblending of lines from 
both stars. 

\begin{table}
\caption{Estimate of expected crowding of the spectra. See text.}
\begin{tabular}{lrrrr}
                      &\multicolumn{2}{c}{$V<20$}&\multicolumn{2}{c}{$V<17$}\\ 
		      &$|b|<20^o$&$|b|>20^o$&$|b|<20^o$&$|b|>20^o$\\ \hline
$n$ [stars/ deg.$^2$]& 61000    & 5600     & 6100     & 1200 \\
aver.\ separation    & 15"      & 48"      & 46"      & 104" \\
free length ($l$)    & 212"     & 2310"    & 2120"    & 10800"\\ \hline
\end{tabular}
\end{table}

The Galaxy model (Torra et al.\ 1999) using Hakkila et al. (1997) 
extinction law predicts an average  surface density $n$ of stars (Table 3).
This can be translated into an average stellar separation ($n^{-0.5}$).
Two spectra are blended if their spectral tracings are separated
by less than the spatial extent $s$ of a CCD pixel, i.e. 1 arcsec.
This happens if the length of the spectral tracing is larger than 
the free length $l = (n s)^{-1}$. The length of the tracing depends on the 
dispersion and equals  1000 arcsec for 0.25 \AA/pix,  500 arcsec for 
0.5 \AA/pix and  333 arcsec for 0.75 \AA/pix. So Table 3 demonstrates 
that spectral superposition will be common close to the galactic plane, 
but not away from it. 
 
Light of contaminating stars 
will be spread over many pixels, so the zodiacal light will usually remain 
the dominant background signal. Stellar background features stellar lines 
and this could influence the radial velocity accuracy. Still, extremely 
accurate astrometry and photometry of contaminating stars will supply 
enough information to handle the problem. In the worst cases the 
situation will be similar to disentangling the two spectra in a spectroscopic
binary. But with an extra benefit that spectral overlap varies from 
transit to transit. So, even if two spectra are badly superimposed,
the same two stars will be recorded separately during the next transit.

\section{Conclusions}

We studied performance of a slitless spectrograph observing in a 
250~\AA\  window centered at  8615~\AA\  that will be used aboard GAIA.
The same spectral range is accessible also from the ground, 
but cannot be much expanded, due to bracketing telluric absorptions. 
So the results are relevant also for planning of ground based instruments. 

The dominant parameter that can still be changed is the spectral resolution. 
We used realistic estimates of zodiacal light background and spectrograph 
parameters to derive expected accuracy of the radial velocity determination. 
We showed that errors below 2~km/s are achievable for bright enough 
stars if a high dispersion is chosen. This result could be still improved 
with correlation methods that use software spectral masks 
(Baranne et al.\ 1979, Queloz 1995). 

The results show that it is worth to observe very faint targets, down to 
magnitude 19  even though the expected radial velocity errors can be 
well above 10 km~s$^{-1}$. 
Such stars are numerous. One may expect a ratio of 
$\sim$ 10:1 in the recorded number of targets between V=17 and 19 compared to 
stars between V=14 and 16. For such stars the best results are obtained with 
0.5~\AA/pix or 0.75~\AA/pix dispersion. But for brighter 
stars a high dispersion (0.25~\AA/pix) is always a preferred choice 
and is critical in order to extract relevant information for non-trivial 
cases, as spectroscopic binaries, variable stars and peculiar stars. 
  
It remains to be determined if telemetry constraints permit 
to use the dispersions of 0.25 or 0.5 \AA/pix. Still 
it is clear that such a choice has a strong support considering the 
astrophysical information retrievable from the data. 
\smallskip \\ 
{\bf Acknowledgement.} The author wishes to thank U.\ Munari for 
initial motivation and many fruitful discussions and to F.\ Castelli for an  
introduction to the art of Kurucz models. The referee, D.\ Katz, is 
acknowledged for useful comments which improved the quality of the 
paper. This work was supported by a grant from the Slovenian Ministry of 
Education, Science and Sport.

\end{document}